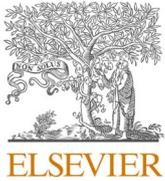

Contents lists available at ScienceDirect

## Photoacoustics

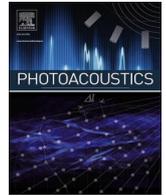

# Deep image prior for undersampling high-speed photoacoustic microscopy

Tri Vu [a,*], Anthony DiSpirito III [a], Daiwei Li [a], Zixuan Wang [c], Xiaoyi Zhu [a], Maomao Chen [a], Laiming Jiang [d], Dong Zhang [b], Jianwen Luo [b], Yu Shrike Zhang [c], Qifa Zhou [d], Roarke Horstmeyer [e], Junjie Yao [a]

[a] *Photoacoustic Imaging Lab, Duke University, Durham, NC, 27708, USA*
[b] *Department of Biomedical Engineering, Tsinghua University, Beijing, 100084, China*
[c] *Division of Engineering in Medicine, Department of Medicine, Brigham and Women's Hospital, Harvard Medical School, Cambridge, MA, 02139, USA*
[d] *Department of Biomedical Engineering and USC Roski Eye Institute, University of Southern California, Los Angeles, CA, 90089, USA*
[e] *Computational Optics Lab, Duke University, Durham, NC, 27708, USA*

## ARTICLE INFO

## ABSTRACT

Photoacoustic microscopy (PAM) is an emerging imaging method combining light and sound. However, limited by the laser's repetition rate, state-of-the-art high-speed PAM technology often sacrifices spatial sampling density (*i.e.*, undersampling) for increased imaging speed over a large field-of-view. Deep learning (DL) methods have recently been used to improve sparsely sampled PAM images; however, these methods often require time-consuming pre-training and large training dataset with ground truth. Here, we propose the use of deep image prior (DIP) to improve the image quality of undersampled PAM images. Unlike other DL approaches, DIP requires neither pre-training nor fully-sampled ground truth, enabling its flexible and fast implementation on various imaging targets. Our results have demonstrated substantial improvement in PAM images with as few as 1.4 % of the fully sampled pixels on high-speed PAM. Our approach outperforms interpolation, is competitive with pre-trained supervised DL method, and is readily translated to other high-speed, undersampling imaging modalities.

## 1. Introduction

With unique optical absorption contrast, photoacoustic imaging (PAI) has been increasingly used for preclinical and clinical applications [1,2]. PAI employs short-pulsed laser excitation and detects the resultant ultrasonic pressure waves. Photoacoustic microscopy (PAM) is a major implementation of PAI that uses tightly or weakly focused optical excitation and acoustic detection. Different from confocal and two-photon microscopy, PAM can achieve an imaging depth beyond the optical diffusion limit with high spatial resolution. As a result, PAM has been a powerful tool for studying small animal models *in vivo*, particularly for imaging microvasculature using hemoglobin as the endogenous contrast [3,4].

Similar to other microscopic imaging modalities, PAM relies on point-by-point scanning to acquire volumetric images over a large field of view (FOV). In order to achieve the desired spatial resolution, the raster scanning needs to satisfy the Nyquist-Shannon sampling theorem, which requires a scanning step size of at most half of the desired spatial resolution. For example, as most PAM studies target blood vessels with a diameter of at least 10 μm [2,5,6], a scanning step size of 5 μm is often satisfactory [7–9]. However, the high spatial sampling density often leads to a low imaging speed of traditional PAM systems with mechanical scanning [8–11]. In recent years, state-of-the-art fast PAM systems with fast scanners, such as Galvo mirror, micro-electromechanical mirror, and polygon scanners, have achieved substantially improved scanning speeds [12–16]. Nevertheless, unlike traditional PAM systems relying on slow mechanical scanning, the imaging speed of fast PAM systems is often limited by the laser's pulse repetition rate. For example, we recently reported a polygon-scanner-based PAM system, which has reached a B-scan rate along the fast-scanning axis as high as 2000 Hz, over a 10 mm scanning range [17]. However, with a laser's pulse repetition rate of 800 kHz, to satisfy the Nyquist sampling, the B-scan rate is limited to only 200 Hz, much lower than the maximum capacity of the system. One way to increase the imaging speed is to enlarge the scanning step size. Sparse sampling has often become a necessary compromise for imaging speed in fast PAM, which results in degraded image quality including lower resolution and spatial aliasing.

* Corresponding author.
  *E-mail address:* tqv@duke.edu (T. Vu).

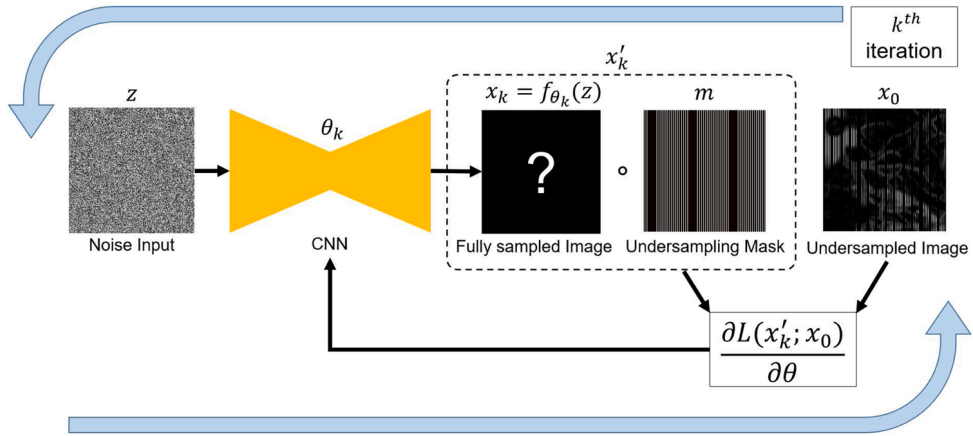

**Fig. 1.** Schematics of the iterative DIP optimization. At the $k^{th}$ iteration, the gradient of $x_k'$ and $x_0$ w.r.t $\theta$ is calculated and back-propagated to the encoder-decoder CNN model.

The advances in deep learning (DL) have been progressively incorporated with PAI. Over the past few years, researchers have primarily focused on applying DL in photoacoustic computed tomography (PACT) for artifact removal, target identification, and sparse sampling [18–26]. Recently, DL has been used to reduce the laser pulse energy [27] and improve the undersampling in PAM [28,29]. For the undersampling issue, previous applications on *in vivo* PAM data, with a structural similarity index (SSIM) of up to 0.92 [28,29]. However, the previous DL methods typically depend on training convolutional neural networks (CNN) with a large training dataset as the ground truth. Collecting the *in vivo* training dataset is time- and resource-intensive, if available at all. Additionally, pre-trained models are inclined to be biased toward learned features of the training datasets, and thus likely to fail on unfamiliar input images [25,30,31]. Furthermore, adding new training samples requires model retraining and thus takes a longer time. A DL model that does not require prior training on large ground-truth data is highly desired for improved generalization.

The emerging deep image prior (DIP) method requires neither pre-training nor labeled ground truth data [32]. The DIP network acts as a parametrization, which is solved by iterative optimization [32]. Taking only a noise image as the input, the DIP model outputs the restored image. Following a known distortion operator (*e.g.*, downsampling pattern), the output is then compared to the target image (*e.g.*, downsampled PAM image) in the objective function for backpropagation and optimization. Although traditional supervised DL methods require pre-training with ground truth, the DIP model requires only the target image and the distortion operator. The DIP model learns features of the distorted images from the applied convolutional filters, and then imposes these implicit traits (*i.e.*, prior) on the restored output [32]. The DIP method has shown great promise for improving medical imaging, such as in-painting, denoising, and artifact removal in PET, diffraction tomography, and scanning electron microscopy [30,32–35]. Because PAM often lacks a large *in vivo* training dataset, we expect that PAM may also benefit from non-pretrained DIP.

In this study, we have developed a new method using a non-pretrained DIP model to improve undersampled PAM images and increase the imaging speed of fast PAM systems. Our DIP model iteratively seeks an optimized, fully sampled image that approximates the undersampled image, given a known downsampling mask. Our approach eliminates the need for pairwise training, enabling its generalization on various imaging results. Our model was tested on mouse brain vasculature images and bioprinted gel images, first on a slow-scanning PAM system and then a fast-scanning PAM system. The performance of our DIP method was also thoroughly compared with (non)linear interpolation methods and a state-of-the-art pre-trained DL model. The DIP model has shown promising performance on both phantom and *in vivo* results without using any prior training data. To the best of our knowledge, this is the first application of DIP on PAM, which will pave the way for further studies in other PAI technologies without training data.

## 2. Methods

### 2.1. Deep image prior (DIP) for undersampling problem

To formulate a DIP model, we first frame the problem as a typical image restoration task via optimization. Given a distorted image $x_0$, we try to find its originally undistorted image $x$ of a width $W$ and height $H$ through an energy minimization process:

$$x^* = \underset{x}{\operatorname{argmin}} \{L(d(x); x_0) + R(x)\} \tag{1}$$

in which $L(x; x_0)$ is our data-fidelity term, $d(.)$ is the distorting operator, and $R(x)$ is the regularization term. $R(x)$ serves as the embedded prior information about the image. For example, total variation can be used as $R(x)$ to enforce smoothness. The choice of $R(x)$ is critically important to find the optimal solution $x^*$. Various groups have investigated the use of CNN as $R(x)$ [36–38], especially for artifact removal [23]. However, these methods cannot account for $L(d(x); x_0)$ and thus require a large number of training pairs to optimize the model. To get rid of $R(x)$ and avoid searching for $x$ explicitly, one can find $\theta$ for the surjective function $g$, such that Eq. (1) becomes:

$$\theta^* = \underset{\theta}{\operatorname{argmin}} \{L(d(g(\theta)); x_0) + R(g(\theta))\} \tag{2}$$

A correct parametrization of $x$ by $g(\theta)$ can express the explicit prior itself, thus eliminating the prior term. Such parametrization can be represented by a neural network $f_\theta(z)$ with noise input $z$ (following uniform random distribution between 0 and 1) and weight $\theta$:

$$\theta^* = \underset{\theta}{\operatorname{argmin}} L(d(f_\theta(z)); x_0) \tag{3}$$

Eq. (3) can be rewritten as a non-linear least squares solution as:

$$\theta^* = \underset{\theta}{\operatorname{argmin}} \|d(f_\theta(z)) - x_0\|^2 \tag{4}$$

In our undersampling problem, the distorting operator is simply a binary undersampling mask, resulting in the following operator:

$$d(f_\theta(z)) = f_\theta(z) \circ m \tag{5}$$

where $\circ$ is the Hadamard product and $m \in \{0, 1\}^{H \times W}$ is the binary mask of the undersampling pattern, in which "1" denotes sampled pixels and '0' denotes skipped pixels. Substituting $d(.)$ in Eq. (5) to Eq. (4), we

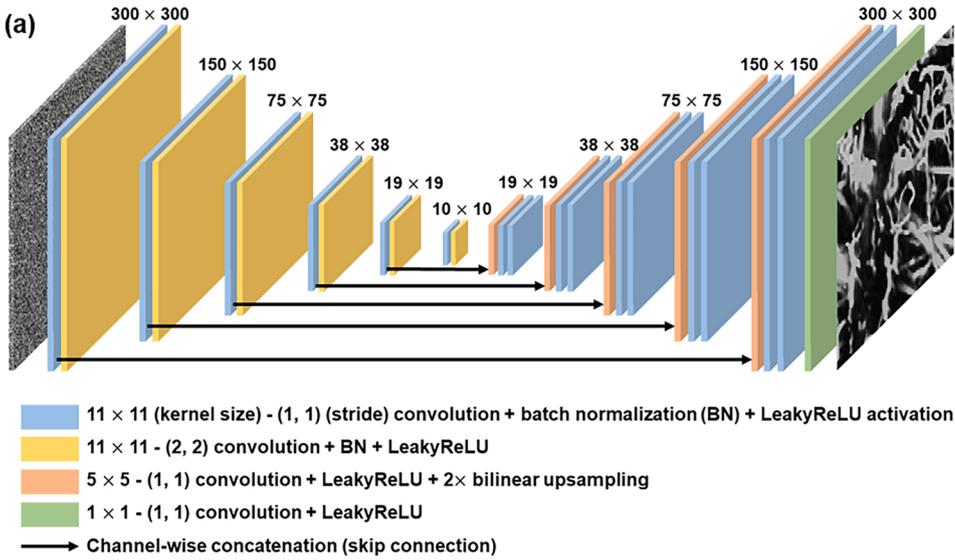

**Fig. 2.** The CNN model used in our DIP model. (a) Network architecture of the modified U-Net. The input is 32-channel $300 \times 300$ random noise, and the output is single-channel recovered PAM image. All other layers have 64 filters. (b) Output image with checkerboard artifact using model with transposed convolution. Our DIP model, with standard convolution and upsampling, eliminates this artifact. Scale bar: 0.5 mm. (c) Training stability is improved by AMSGrad, while Adam has exploding gradient at ~4,000th iteration. (d) SSIM with different $\sigma_z'$ values for our study. All SSIMs are averaged over 15 samples. Error bar: standard deviation.

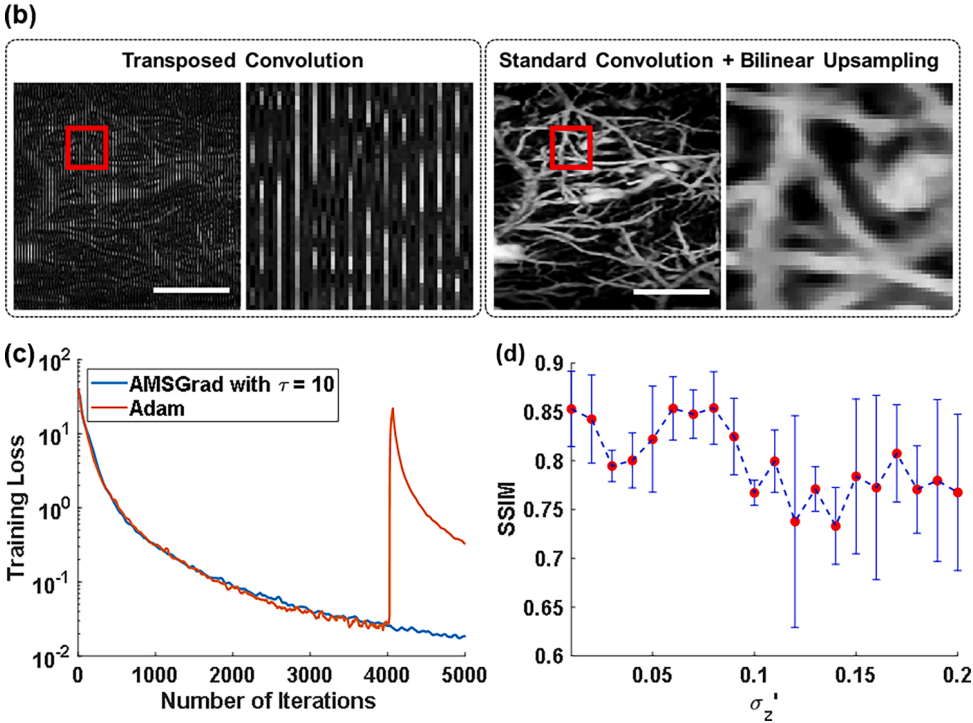

obtain the final formulation of DIP optimization for undersampling restoration:

$$\theta^* = \underset{\theta}{\mathrm{argmin}} \|x' - x_0\|^2 \text{ with } x' = f_\theta(z) * m \quad (6)$$

Generally, instead of directly optimizing the fully sampled image $x$, which requires a large number of training pairs, we iteratively update the weight ($\theta$) of the CNN, which outputs $x = f_\theta(z)$. This output, multiplied by the sampling mask, is expected to match the undersampled image $x_0$. The DIP optimization scheme is shown in Fig. 1.

In this study, we use the noise-based regularization in the input, as recommended by [32]. Instead of using a fixed noise input, we add to $z_0$ an additional perturbing noise term $z'$ with a Gaussian distribution. $z'$ has a zero mean and an experimentally determined standard deviation $\sigma_z = 0.07$ (Fig. 2(d)) [32]. This perturbing noise helps prevent overfitting and results in a more generalized solution [39,40]:

$$z = z_0 + z' \quad (7)$$

## 2.2. Network architecture and training

Our CNN for the deep prior employs U-Net with some modifications [41]. U-Net is a popular network architecture for DL, especially for medical imaging applications. Its encoder-decoder framework effectively extracts features in the compression (encoder/downsampling) path and reconstructs the target from those features in the decompression (decoder/upsampling) path. There are two reasons for using a U-Net architecture in DIP. First, with a random weight and noise input, U-Net can reconstruct spatial structures with self-similarities, regardless of the stochastic inner parameters [32]. Thus, it is possible that, after being exposed to the distorted image, U-Net can apply certain similarities learned from the existing structures to the rest of the recovered image. The second reason lies in the inter-layer concatenations of U-Net. By concatenating channels from deep layers, U-Net can pass high-level features to the upsampling scheme and create self-similarities on the output at different scales [32].

The architecture of our modified U-Net is shown in Fig. 2(a). In

 

**Table 1**

Comparison of training and inferring time between DIP and FD U-Net. Consistent with the reported FD U-Net in [29], the input of DIP is 128 × 128 pixels.

|  | DIP | FD U-Net [29] |
|---|---|---|
| Training time | 2.60 min | 8 h |
| Inference time | – | 46.0 ms |

general, it contains the main characteristics of U-Net. The modifications are mainly in the inner blocks. First, we keep the same number of filters (64) for all the layers to decrease the number of model parameters. In the original U-Net model, the number of filters increases in the down-sampling path to extract the features and decreases in the upsampling path to localize the features for semantic segmentation [41]. In our study, we instead use the same number of layers in order to extract and reconstruct high-resolution features at all spatial levels. In addition to this channel reduction, we also incorporate a large kernel size (11 × 11) for the network. Large receptive kernel can more readily draw from the information of nearby pixels in the undersampled image. In addition, we avoid using a transposed (sub-pixel) convolutional layer in the decoder path, which is known to introduce severe checkerboard artifacts due to the deconvolution overlap and random initialization (Fig. 2(b)) [42–44]. Instead, as suggested by [42,44], we split this upsampling process into standard 2D spatial convolution and bilinear interpolation, thus eliminating the aforementioned artifacts (Fig. 2(b)).

### 2.3. Model training and evaluation

The modified U-Net was trained with the following general settings. The optimizer was AMSGrad with a clip value of $\tau = 10$, which has shown significant improvement in training stability compared to Adam (Fig. 2(c)). We trained the model for 5,000 iterations and used mean-squared error (MSE) loss for all experiments. All convolutional layers were followed by Leaky ReLU activation ($\alpha = 0.2$) instead of ReLU to help combat the vanishing gradient [45]. The model was trained on a 64-bit workstation with an NVIDIA RTX 2080 Ti GPU and an Intel Core i9-9900 K, using Python v3.6.1 and Keras 2.2.5 with Tensorflow backend. Optimizing a 300 × 300 pixel image took approximately 14.3 min. Detailed training parameters are shared for public access at https://github.com/trivu169/deep-prior-pam.

The model performance was evaluated using the SSIM and peak signal-to-noise ratio (PSNR) on the testing dataset. These metrics represent both global and local information of the reconstructed fully-sampled images [25]. SSIM and PSNR were also computed for bilinear, bicubic, and lanczos (8 × 8 kernel) interpolation as well as our recently published pre-trained FD U-Net [29]. In FD U-Net, each block at a certain spatial level has channel-wise concatenation of all convolutional layers as the output, which forms "collective knowledge" that is passed to the next block [29,46]. A comparison of training and inferring time between DIP and FD U-Net is summarized in Table 1. Please note that, in DIP, since each sample needs iterative training, the training time also includes the inferring time. Statistical $p$-values were calculated to compare the performance of DIP with the interpolation methods and FD U-Net, using ANOVA with post-hoc Tukey's HSD test [47]. The testing dataset contained 37 vascular images with 300 × 300 pixels. A total of five undersampling patterns was used for evaluation, as delineated in the next section. To accommodate large images (i.e., > 300 × 300 pixels), we used a *model patchwork algorithm* to avoid border artifacts [29]. Given enough GPU memory, it is also possible to process the whole image at once without patching.

### 2.4. Data acquisition and undersampling mask construction

Two PAM systems were used in this study. The first was the traditional slow-scanning PAM system using motorized stages [9,48], with an effective spatial resolution of 10 μm, a scanning step size of 5 μm, and a

**Table 2**

Comparison of training time for different undersampling patterns with an input of 300 × 300 pixels (mean ± standard deviation).

| Undersampling ratio | [4,1] | [5,1] | [2,5] | [6,1] |
|---|---|---|---|---|
| Training time (mins) | 14.6 ± 0.06 | 14.8 ± 0.03 | 14.7 ± 0.17 | 13.8 ± 0.14 |
| Undersampling ratio | [7,3] | [10,5] | [6,12] | [8,14] |
| Training time (mins) | 14.8 ± 0.03 | 14.7 ± 0.04 | 13.6 ± 0.02 | 13.7 ± 0.02 |

B-scan rate of 1 Hz. Thus, it takes ~16.7 min to complete 1000 B-scans. The slow-scanning PAM system provided fully-sampled images as the testing dataset. The second system was the recently published fast-scanning PAM system using a polygon scanner, with a spatial resolution of 10 μm and a B-scan rate of 1 kHz, improving the scanning speed by 1000 times compared to slow-scanning system. Limited by the laser's repetition rate, the fast-scanning PAM provided only under-sampled images.

All experimental protocols were approved by the Institutional Animal Care and Use Committee (IACUC) of Duke University. We imaged the mouse brain and ear vasculature *in vivo*, as previously demonstrated [2,6,49–52]. In order to demonstrate the DIP model's generalization, we also imaged a bioprinted hydrogel sample [53]. All images were acquired at 532 nm and presented as the maximum amplitude projection images.

For the slow-scanning PAM, similar to [29], we artificially generated the undersampled images from the fully sampled imaged. We first denoted $[S_x, S_y]$ as the undersampling ratio in the $x$ and $y$ axis, which was used to construct a binary undersampling mask of "0" and "1". The effective pixels are defined as the pixels in the undersampled image exacted from the fully sampled image. For example, the undersampling ratio [5,1] provides 20 % effective pixels. The mask was then pixel-wise multiplied with the fully sampled image to create the undersampled image. Eight representative undersampling ratios were used in this study for slow-scanning PAM: [8,14], [6,12], [10,5], [7,3], [6,1], [2,5], [5,1], and [4,1]. One undersampling ratio of [6,12] was used for fast-scanning PAM, as determined by the system's maximum speed. The averaged training time for each pattern is shown in Table 2.

### 3. Results

#### 3.1. DIP on in vivo vascular data

Representative outputs from all methods are shown in Figs. 3 and 4. The improvements by DIP are more significant for high undersampling ratios (<20 % effective pixels), as shown in Fig. 3. Particularly, even at [7,3] and [10,5], blood vessels recovered by DIP remain relatively continuous and smooth. This is a key attribute of DIP, which does not learn the vascular continuity from any ground truth. In contrast, interpolation-based methods result in discontinuous and aliased vessels, as shown in Fig. 3. The staircase artifacts cause severe intensity fluctuation and even shift the vessel positions (Fig. 3(e, f)). From the vessel profiles, DIP is able to accurately reconstruct the correct signal intensities along different directions. As expected, the pre-trained FD U-Net yields slightly better performance than DIP. FD U-Net results show sharper and more continuous vessels than the DIP results (Fig. 3(c)). At low-to-moderate undersampling ratios (e.g., [4,1]), the difference between all methods is less obvious (Fig. 4). Nevertheless, both FD U-Net and DIP keep the key improvements on vessel continuity and connectivity at [5,1] and [6,1] (Fig. 4(d, e)). However, it is important to note that, when a structure is completely skipped by the undersampling scheme, especially at a high sparsity ratio of [10,5] and more, it cannot be recovered by neither DIP nor other methods, such as the vertical vessel denoted by the dashed ellipse in Fig. 3(d).

Fig. 5 shows an illustration of the DIP optimization process with an undersampling ratio of [7,3]. Note that the model evolves from an arbitrary output (0$^{\text{th}}$ iteration) to a meaningful representation by the $1000^{th}$ iteration (Fig. 5(a)). The evolution is consistent with the learning curves in Fig. 5(b, c), with a sharp increase in SSIM and PSNR after the first 1000–1500 iterations. Further optimization up to 3000 iterations mainly recovers finer details and smaller vessels. There is no obvious improvement after 3000 iterations in SSIM and PSNR (Fig. 5(b, c)). Therefore, it is possible to reduce the training time by 40 % from 14.3 min to 8.6 min, without sacrificing the reconstructed image quality.

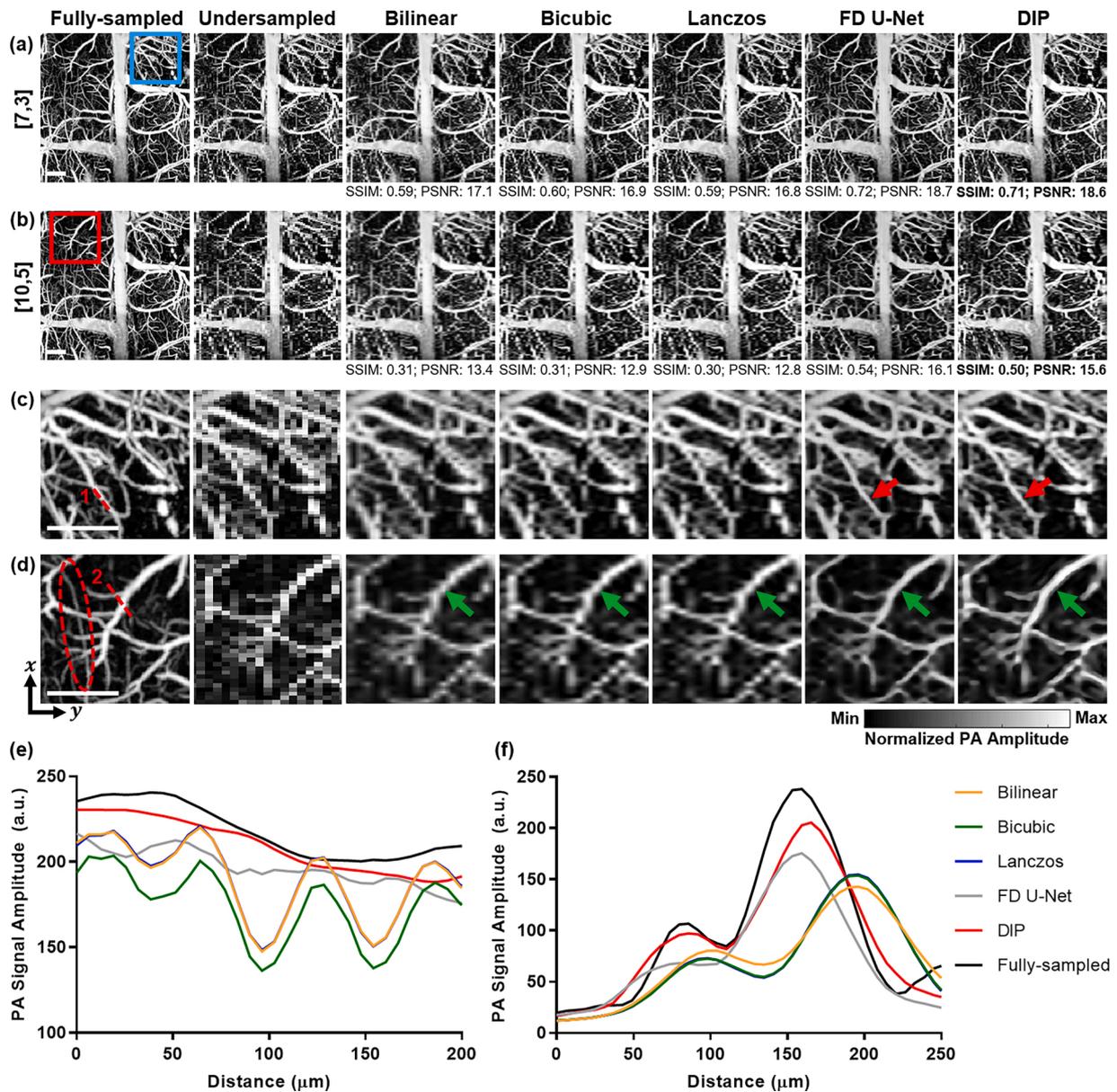

**Fig. 3.** DIP of representative undersampled PAM images with high undersampling ratios. (a-b) Representative $600 \times 600$ pixel output from DIP and other methods with undersampling ratios of [7,3] and [10,5], respectively. Lanczos is the representative interpolation method. (c-d) Close-up regions of the blue and red boxes in (a-b), respectively. The arrows indicate broken or jagged vessels in interpolation results, but not in the DL results. The red dashed ellipse denotes the vessel missed by all methods. (e-f) Image pixel intensity profiles along the line 1 and 2 in.(c-d), respectively. Scale bar: 0.5 mm.

After validating the DIP model using the slow-scanning PAM with full-sampled ground truth, we further applied the DIP model to the mouse vasculature data acquired by our recently developed fast-scanning PAM, with an undersampling ratio of [6,12] (Fig. 6). For this extremely sparse sampling (only 1.4 % effective pixels), we can further test the performance of the DIP model for significant improvement in imaging speed. Similar to the results by slow-scanning PAM, the severe vasculature aliasing results in distorted and discontinuous vessel profiles by the interpolation methods, shown in Fig. 6. By contrast, the results by

DIP and FD U-Net show improved vessel continuity and edge smoothness. The pre-trained FD U-Net, however, has checkerboard artifacts due to either residual scanning effects or transposed convolution (Fig. 6(c)). DIP is visually free of unidentified artifacts, and has the best vessel connectivity and least edge jags. Interestingly, DIP can better resolve closely-located vessels while the interpolation methods and FD U-Net fail (yellow arrows in Fig. 6(d)). For the fast-scanning PAM, the undersampling ratio of [6,12] can lead to a 72-fold improvement in imaging speed, which is mainly limited by either the laser's repetition rate or the number of effective pixels.

### 3.2. Quantification of DIP performance

The performance of different methods was quantified with SSIM and PSNR (Fig. 7). FD U-Net has the best quantitative performance, followed by DIP. For SSIM, FD U-Net outperforms DIP ($0.756 \pm 0.096$ vs. $0.693 \pm 0.109$) with the [10,5] undersampling ratio (Fig. 7(a)).

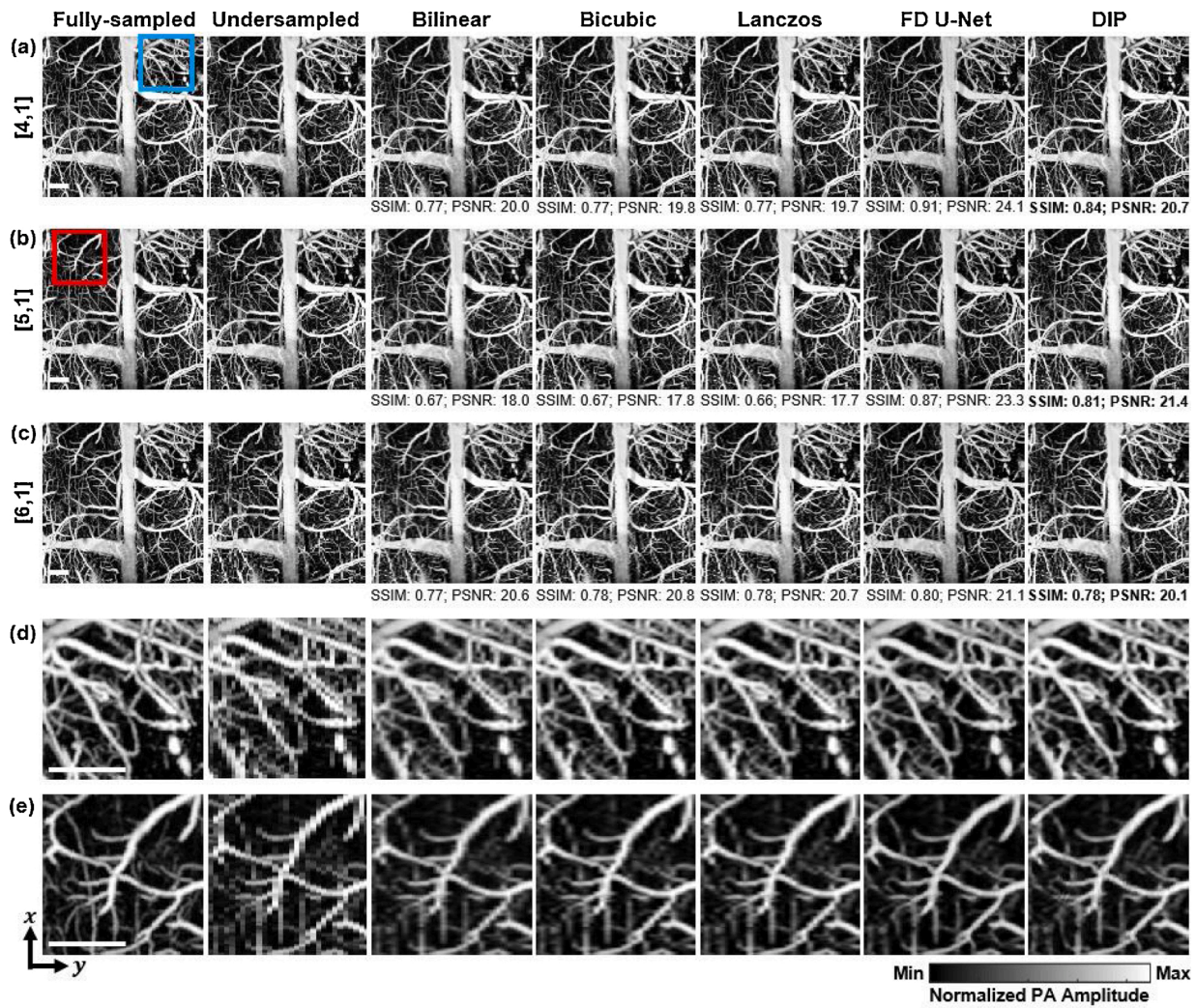

**Fig. 4.** DIP of representative undersampled PAM images with low-to-moderate undersampling ratios. (a-c) Representative $600 \times 600$ pixel output from DIP and other methods with undersampling ratios of [4,1] [5,1], and [6,1]. (d-e) The close-up regions indicated by the blue and red boxes with [5,1] and [6,1] undersampling ratios. Scale bar: 0.5 mm.

However, the difference reduces as the undersampling ratio decreases to [6,1] ($0.910 \pm 0.039$ vs. $0.903 \pm 0.047$). For PSNR, the differences between DIP and FD U-Net are similar for all undersampling ratios, likely because PSNR and the loss function of the DIP model both depend on MSE. DIP outperforms all other methods at [6,1] with the highest PSNR of $29.71 \pm 4.23$ dB (Fig. 7(b)).

The statistical tests also confirm that DIP outperforms interpolation methods, except for the low undersampling ratio [4,1] (Fig. 7(c, d)). In fact, $p$-values show that DIP and FD U-Net are indeed comparable for both PSNR and SSIM for almost all undersampling patterns, except [4,1]. Interestingly, at this ratio, both interpolation and FD U-Net have significantly larger SSIM than DIP, but there are no significant differences for PSNR. This result suggests that DIP is more advantageous for high sparsity (<25 % effective pixels), and such advantage diminishes with more effective pixels. In addition, even though there is no theoretical limit for DIP, Fig. 7(a, b) suggest a steady decrease in SSIM and PSNR with increasing sampling sparsity. At extreme undersampling ratios of [6,12] and [8,14] with only 1.4 % and 0.9 % effective pixels respective, $p$-values show that DIP does not have significantly higher SSIM and PSNR than other methods. However, the result of the fast-scanning PAM data shown in Fig. 6 demonstrates that DIP still offers much improvement in vascular continuity.

### 3.3. DIP on non-vascular data

In order to test DIP's generalization on non-vascular images, we also imaged a bioprinted sample of Duke Chapel using the slow-scanning PAM. The representative outputs of a [7,3] undersampling ratio are shown in Fig. 8. The results are consistent with the *in vivo* vascular results. The interpolation results show blurry and jagged edges. The FD U-Net output suffers from the checkerboard artifacts. The DIP image, on the other hand, has shown clear and smooth edges, and no artifacts. This non-vascular phantom result demonstrates that DIP may also be applied to novel input images without the need for pre-training. Both DL methods do not enforce any vessel-like features on this phantom data, asserting their generalization to various input structures.

### 4. Discussion

One of the critical missions in improving PAM technology is to increase the imaging speed for time-sensitive applications. Unlike traditional PAM systems, which rely on slow mechanical scanning, the imaging speed of state-of-the-art, high-speed PAM systems is often limited by the laser's repetition rate or the total number of effective pixels. Undersampling is often a necessary compromise when imaging speed needs to be increased. Approximately, the imaging speed of fast-scanning PAM is inversely proportional to the number of effective pixels. Therefore, reducing the number of pixels through undersampling will lead to the same percentage increase in the imaging speed.

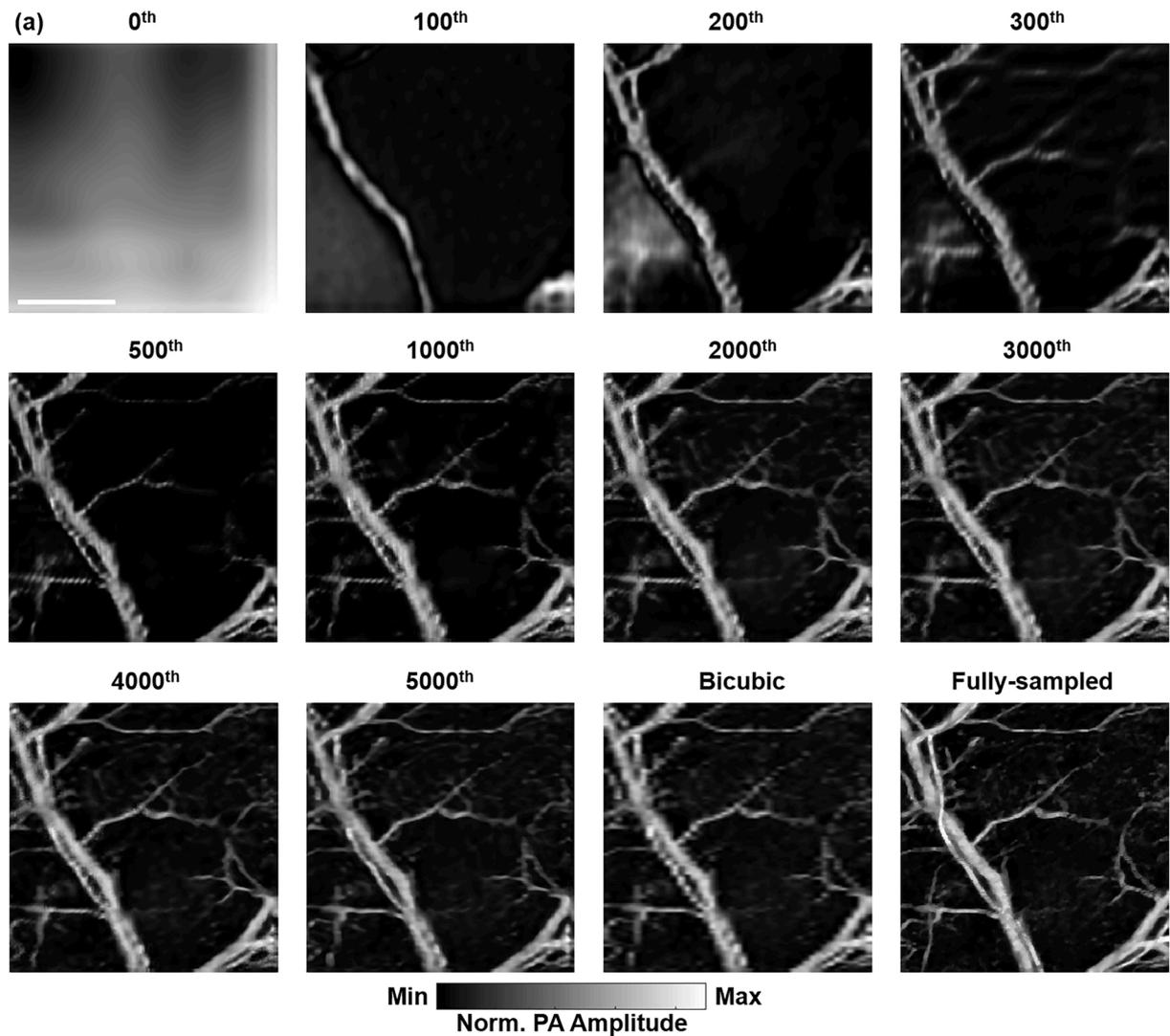

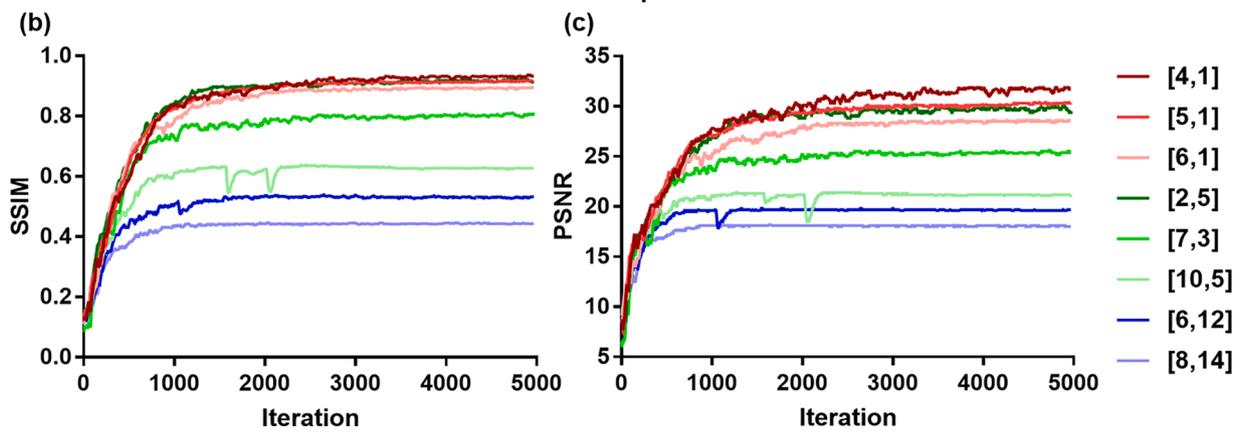

**Fig. 5.** Evolution of the DIP optimization process. (a) Output image at incrementing iterations. The image is undersampled at [7,3]. The bicubic interpolation and fully-sampled image are also presented. Scale bar: 0.5 mm. (b-c) Averaged SSIM and PSNR of three random samples as a function of iterations.

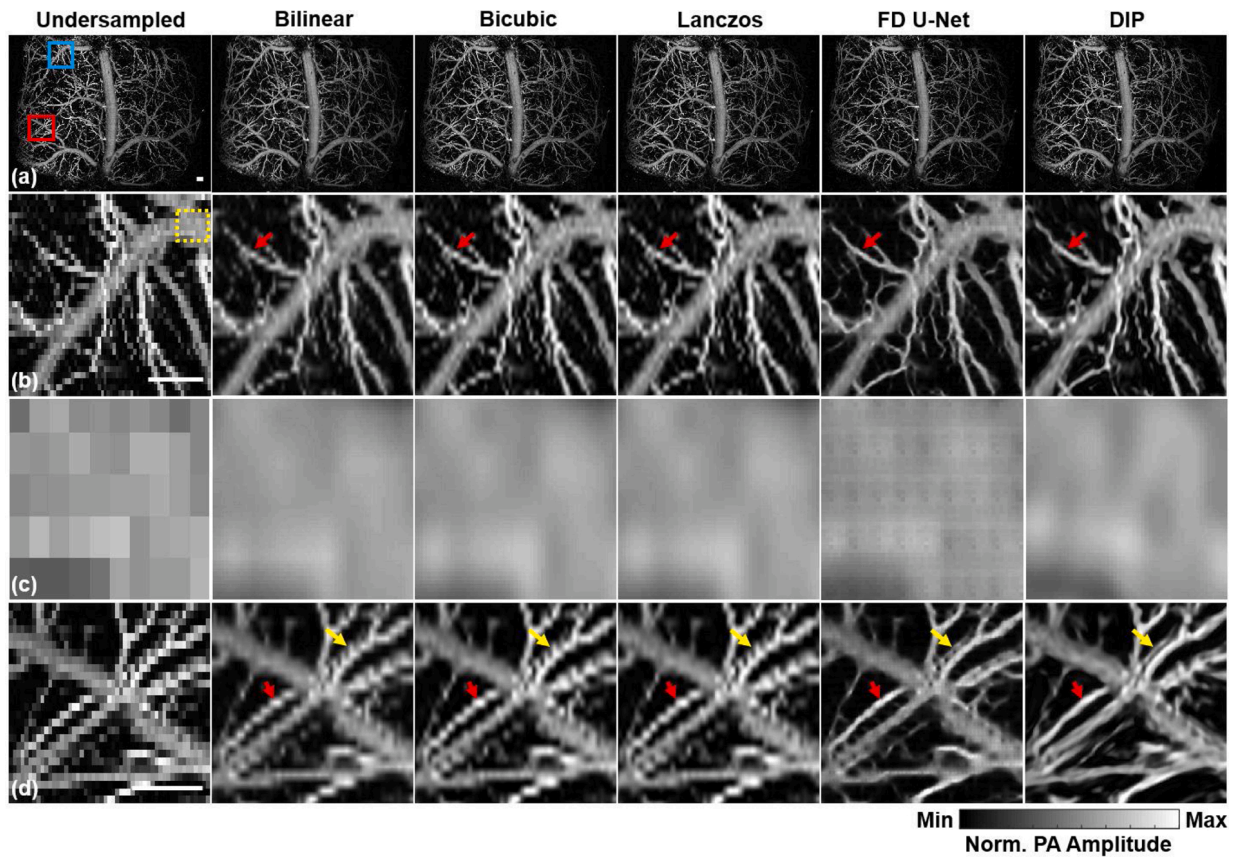

**Fig. 6.** Outputs from DIP and other methods on fast-scanning PAM image with an undersampling ratio of [6,12]. (a) Wide-field mouse brain vasculature images. (b) Close-up region marked by blue box in (a). (c) Close-up areas in the yellow square in (b) showing the checkerboard artifacts by FD U-Net but not by other methods. (d) Close-up region marked by the red box in (a). The red arrows denote disconnected and jagged vessels that are not observed in the DL methods. Yellow arrows indicate better-resolved vessels by DIP compared to FD U-Net. Scale bar: 125 μm.

Despite the lack of pre-training with fully-sampled ground truth data, DIP manages to correct undersampled images with high accuracy and fidelity. The reconstructed vascular images by DIP maintain a high level of connectivity and edge smoothness – key physiological features of blood vessels. By contrast, interpolation-based methods introduce disjointed and blurry vessels. Similar to other pre-trained DL-methods, FD U-Net may generate checkboard artifacts, primarily due to insufficient training, inappropriate network architecture, or residual scanning effects. DIP does not result in any spurious artifacts in either *vascular* or *non-vascular* data.

Quantitative analysis confirms comparable performance between DIP and FD U-Net. DIP can reconstruct sparsely sampled images using only as low as 0.9 % effective pixels, without modifying the system hardware or collecting large-size fully-sampled training data. DIP may lead to imaging speed improvement for PAM and other high-speed imaging modalities with point-by-point scanning. In fact, DIP can easily adjust the undersampling masks to adapt different sampling strategies, such as spiral, Lissajous, or sine-wave scanning. This flexibility is not shared by other pre-trained DL methods that require retraining if the undersampling mask is changed. However, DIP's advantage over interpolations diminishes with either extremely low or high sparsity. Thus, the number of effective pixels that is most beneficial for DIP needs to be optimized for different imaging systems and imaged targets.

Nonetheless, our DIP model can be further improved. The results show that our DIP model's advantage over interpolation diminishes at low undersampling ratios. DIP is thus more suited for highly undersampling applications capturing dynamic information, such as blood flow, neuronal activities, and circulating tumor cells. Similar to other upsampling methods, DIP cannot recover structures that are completely missed by the undersampling scheme. The main disadvantage of our current DIP model is the relatively long reconstruction time with iterative optimization. Optimizing a 300-by-300-pixel image takes approximately 14.3 min, which is impractical for real-time image processing. Our future work will focus on speeding up the optimization of DIP model. Reconstruction time can be shortened by reducing the total number of model parameters, at the cost of image quality. The trade-off can be addressed with transfer learning by integrating upper layers of a pre-trained neural network, such as VGG-Net or FD U-Net, with the DIP model. During DIP optimization, these pre-learned layers will be locked, which would reduce the number of trained parameters and thus the training time. Integrating the pre-learned features may also reduce the optimization complexity of the deep prior and potentially improve image quality. Future works will also aim at using DIP for other PAI implementations. For example, we can apply DIP in photoacoustic computed tomography (PACT) to reduce the image artifacts induced by sparse sampling and limited detection view. Similar to PAM, PACT often lacks ground truth images at the large penetration depth (> 1 cm), and thus, it is challenging to apply pretrained DL models. DIP is a compelling choice for improving PACT applications.

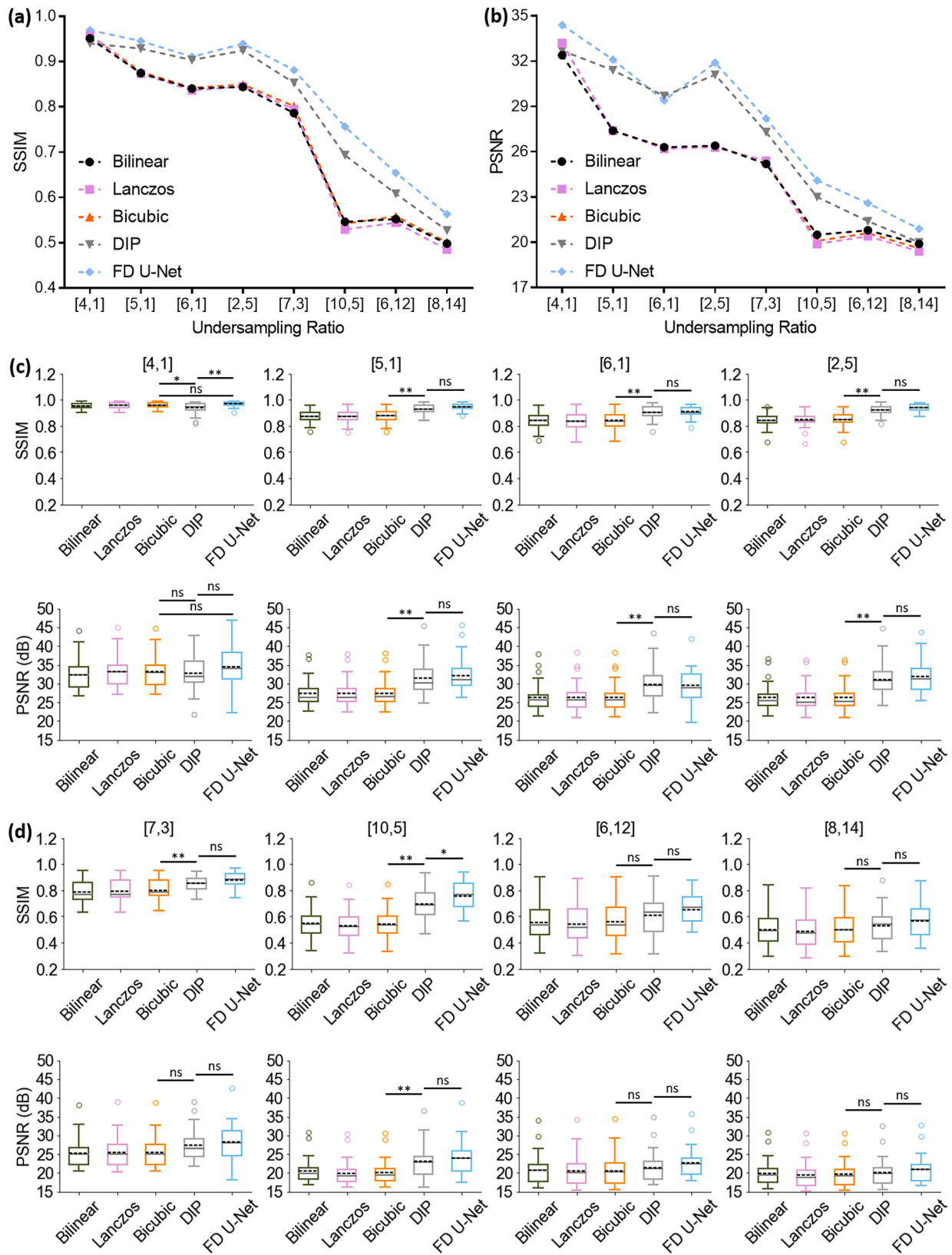

**Fig. 7.** Quantitative comparison of DIP and other methods in SSIM and PSNR at different undersampling ratios. (a-b) Averaged SSIM and PSNR as a function of the undersampling ratios for all methods. (c-d) Boxplots of averaged SSIM and PSNR of low and high undersampling ratios. Most *p* -values between DIP and FD U-Net reveal no significant difference. Meanwhile, DIP shows significant improvement compared with bicubic interpolation, except for [4,1] ratio. Gray line: median; black dashed line: mean; circles: outliers; ns, *p* > 0.05; * *p* ≤ 0.05; ** *p* ≤ 0.01.

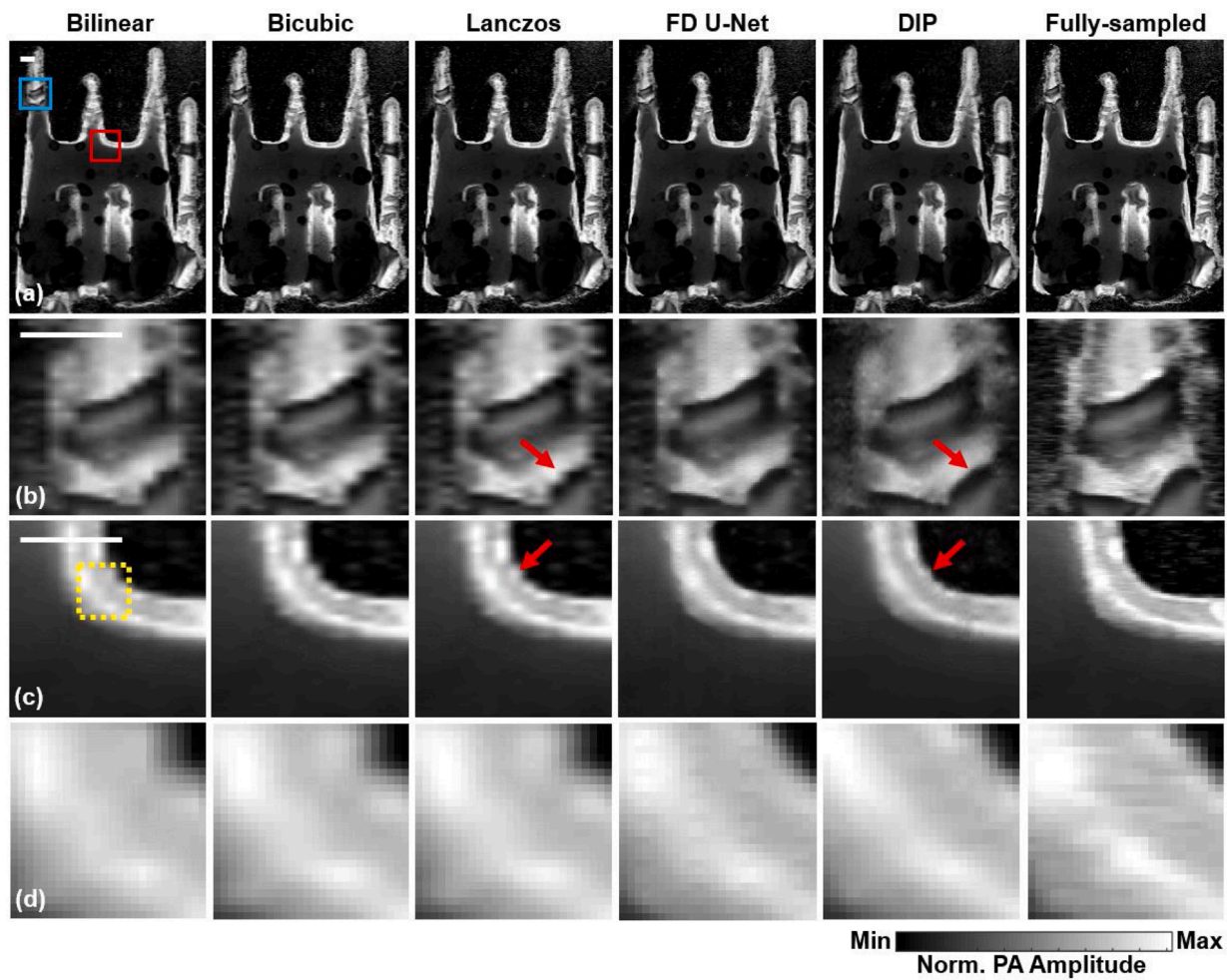

**Fig. 8.** Output of DIP and other methods on a bioprinted sample of the Duke Chapel with a [7,3] undersampling ratio. (a) Wide-field images of the bioprinted sample with (b-c) close-up regions denoted by the blue and red box respectively. Red arrows denote jagged and blurry edges in the interpolation results. (d) Close-up region in the yellow square indicates the checkerboard artifacts from FD U-Net, which is not observed in other methods. Scale bar: 1 mm.

## 5. Conclusion

In this study, we introduce a novel DL method DIP to reconstruct undersampled PAM images for increased imaging speed. DIP does not require pre-training on a large and diverse dataset with paired ground truth. Our experimental results demonstrate the feasibility of applying DIP on improving PAM images of *in vivo* vasculature and bioprinted sample, with as few as 1.4 % effective pixels. DIP outperforms conventional interpolation methods and is comparable with the pre-trained FD U-Net. Our method may provide a practical solution for high-speed PAM applications with severe undersampling.

## Data and code availability

All mouse brain microvasculature datasets used for this study are available upon request. The main code is available on https://github.com/trivu169/deep-prior-pam.

## Declaration of Competing Interest

The authors declare that there are no conflicts of interest.

## Acknowledgements

A portion of our code is adapted from: https://github.com/satoshi-kosugi/DeepImagePrior. We also thank Dr. Caroline Connor and

Lucas Humayun for editing the manuscript. This work was partially supported by National Institutes of Health grants (R01 EB028143, R01 NS111039, RF1 NS115581, R21 EB027304, R21EB027981, R43 CA243822, R43 CA239830, R44 HL138185); American Heart Association Collaborative Sciences Award (18CSA34080277); Chan Zuckerberg Initiative Grant 2020-226178 by Silicon Valley Community Foundation; Duke Institute of Brain Science Incubator Award. We thank Dr. Caroline Connor for editing the manuscript.

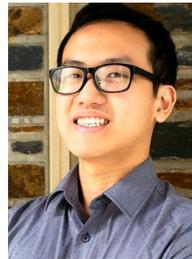

**Tri Vu** received his bachelor's degree in Biomedical Engineering from State University of New York at Buffalo. He is currently a PhD candidate at Department of Biomedical Engineering at Duke University. His research interests are high-speed small-animal photoacoustic imaging systems and photoacoustic image enhancement using deep learning.

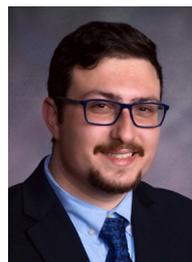

**Anthony DiSpirito** is a biomedical engineering Ph.D. student at Duke University and a Woo Center Fellow at the Duke Institute for Health Innovation. He received his B.S.E. in Biomedical Engineering from Case Western Reserve University with a focus in computational modeling, computer science, and biomedical imaging. He received his M.S. in Biomedical Engineering from Duke University with a focus in artificial intelligence and biomedical imaging. His current research interests center on augmenting biomedical imaging systems using artificial intelligence techniques. As a member of Duke's Photoacoustic Imaging Lab, Anthony has sought to stay at the cutting edge of both deep learning and photoacoustic imaging advances, developing new intelligent imaging systems that embody the growing synthesis between hardware and software.

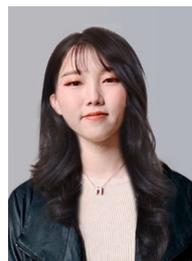

**Daiwei Li** received dual B.S. degrees in Optical Engineering in 2019, one from Changchun University of Science and technology in China, and one from Delaware State University, Dover, DE. She is currently pursuing her Master of Engineering degree in Photonics and Optical Science at Duke University, Durham, NC. She joined the Photoacoustic Imaging Lab (PI-Lab) in the Department of Biomedical Engineering at Duke University in 2019. Her present research interests focus on photoacoustic imaging technologies in the bio-medical field, especially in functional brain imaging using photoacoustic microscopy.

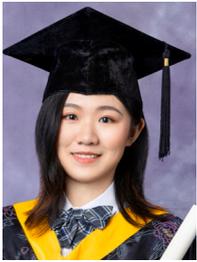

**Zixuan Wang** received her B.S. degree in Biomedical Engineering from Beihang University (China) in 2020. She joined Dr. Yu Shrike Zhang's lab as a research trainee with works on 3D bioprinting, lab-on-a-chip and cancer theranostics at Brigham and Women's Hospital, Harvard Medical School in 2019–2020. Currently, she is a master student at Johns Hopkins University in the Department of Biomedical Engineering, focusing on tissue engineering and precision care medicine.

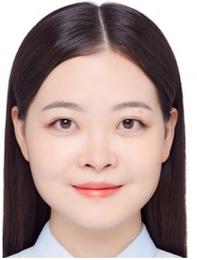

**Xiaoyi Zhu** received her B.S. degree in Biomedical Engineering from Chongqing University in 2014 and Ph.D. degree in Biomedical Engineering from Peking University in 2019. Now, she is a post doctor in Duke University. Her interest focuses on photoacoustic imaging, multimodal imaging and fast scanning imaging system.

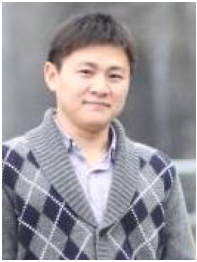

**Maomao Chen** is a postdoctoral research fellow in the BME department at Duke University. He received his Ph.D. degree from the Medical School of Tsinghua University, Beijing, China in 2017. His current research interest focuses on the system development and biological application of photoacoustic imaging.

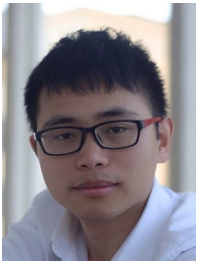

**Dr. Laiming Jiang** is a Postdoctoral Scholar - Research Associate in the Keck School of Medicine at the University of Southern California. He received his Ph.D. degree in Materials Physics and Chemistry from the Department of Materials Science and Engineering, Sichuan University, in 2019. His research work focuses on the development of lead-free piezoelectric materials, ultrasound transducers, energy harvesting, multiscale and multimaterials 3D printing, and bioimplantable devices.

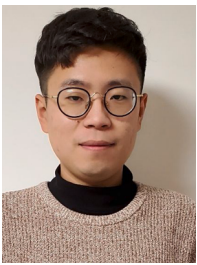

**Dong Zhang** is currently a graduate student at Tsinghua University, China and a visiting scholar at Duke University, USA. His research interests are the development of novel biomedical imaging techniques including photoacoustic imaging and fluorescent moledular imaging.

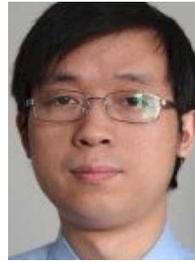

**Jianwen Luo** is an Associate Professor in the Department of Biomedical Engineering, Tsinghua University, Beijing, China. He has authored or coauthored over 150 articles in international journals and 70 conference proceedings articles. His research interests include ultrasound imaging, fluorescence imaging, and photoacoustic imaging. He serves as an Associate Editor for the IEEE Transactions on Ultrasonics, Ferroelectrics, and Frequency Control, an Advisory Editorial Board Member for the Journal of Ultrasound in Medicine, a Technical Committee Member for the IEEE Engineering in Medicine and Biology Society (EMBS) on Biomedical Imaging and Image Processing (BIIP), and a Technical Program Committee

Member for the IEEE International Ultrasonics Symposium (IUS).

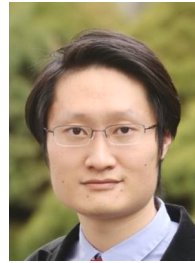

**Dr. Zhang** received a B.Eng. in Biomedical Engineering from Southeast University, China (2008), an M.S. in Biomedical Engineering from Washington University in St. Louis (2011), and a Ph.D. in Biomedical Engineering at Georgia Institute of Technology/Emory University (2013). Dr. Zhang is currently an Assistant Professor at Harvard Medical School and Associate Bioengineer at Brigham and Women's Hospital. Dr. Zhang's research is focused on innovating medical engineering technologies, including bioprinting, organs-on-chips, microfluidics, and bioanalysis, to recreate functional tissues and their biomimetic models towards applications in precision medicine.

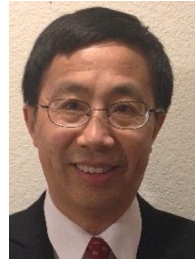

**Qifa Zhou** received his Ph.D. degree from the Department of Electronic Materials and Engineering at Xi'an Jiaotong University of China. He is currently a professor of Biomedical Engineering and Ophthalmology at the University of Southern California. Dr. Zhou is a fellow of the Institute of Electrical and Electronics Engineers (IEEE), the International Society for Optics and Photonics (SPIE), and the American Institute for Medical and Biological Engineering (AIMBE). He has published more than 280 peer-reviewed articles in journals. His research focuses on the development of high-frequency ultrasonic elastography on ocular tissue, intravascular and photoacoustic/OCT imaging for biomedical applications

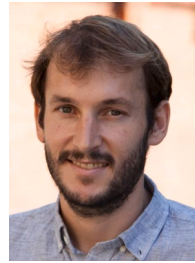

**Roarke Horstmeyer** is an assistant professor of Biomedical Engineering, Electrical and Computer Engineering, and Physics at Duke University. He develops microscopes, cameras and computer algorithms for a wide range of applications, from forming 3D reconstructions of organisms to detecting blood flow and neuronal activity deep within tissue. Most recently, Dr. Horstmeyer was a visiting professor at the University of Erlangen in Germany and an Einstein International Postdoctoral Fellow at Charitè Medical School in Berlin. Prior to his time in Germany, Dr. Horstmeyer earned a PhD from Caltech's EE department (2016), an MS from the MIT Media Lab (2011), and bachelor's degrees in Physics and Japanese from Duke in 2006.

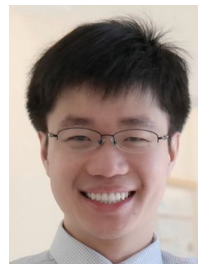

**Junjie Yao** is an assistant professor of Biomedical Engineering at Duke University, and a faculty member of Duke Center for In-Vivo Microscopy, Fitzpatrick Institute for Photonics, and Duke Cancer Institute. He received his B.E. and M.E. degrees at Tsinghua University, and his PhD in Biomedical Engineering at Washington University, St. Louis. His research interest is in photoacoustic tomography technologies in life sciences, especially in functional brain imaging and early cancer detection.